\documentclass[preprint2]{aastex6}

\usepackage{amsmath, amssymb, amsfonts} %Math and such
\usepackage{graphicx} %Allow for figure inclusion
\usepackage{bm} %Bold math
\usepackage{hyperref} %url support
\usepackage{enumitem} %enumerate and itemize formatting
\usepackage[bottom,hang,flushmargin]{footmisc}
\usepackage[none]{hyphenat}
\usepackage{titlesec}
\usepackage{hyphenat} %line break on hyphenated words

\newcommand{\Gpc}{\,\text{Gpc}}
\newcommand{\Jy}{\,\text{Jy}}
\newcommand{\ms}{\,\text{ms}}
\newcommand{\rateUnits}{\,\text{Gpc}^{-3}\,\text{yr}^{-1}}
\newcommand{\skyRate}{\,\text{sky}^{-1}\,\text{day}^{-1}}
\newcommand{\obs}{_\text{obs}}
\newcommand{\nsbh}{_\textsc{nsbh}}

\newcommand{\bns}{_\textsc{bns}}

\newcommand{\frb}{_\textsc{frb}}
\newcommand{\blue}{}

\slugcomment{Published by ApJL}

\begin{document}

\title{Gravitational-Wave Constraints on the Progenitors of Fast Radio Bursts}
\author{Thomas Callister, Jonah Kanner, and Alan Weinstein}
\address{LIGO Laboratory, California Institute of Technology, Pasadena, California 91125, USA; tcallist@caltech.edu}
\date{\today}
\doi{10.3847/2041-8205/825/1/L12}

\begin{abstract}

The nature of fast radio bursts (FRBs) remains enigmatic.
Highly energetic radio pulses of millisecond duration, FRBs are observed with dispersion measures consistent with an extragalactic source.
A variety of models have been proposed to explain their origin.
One popular class of theorized FRB progenitor is the coalescence of compact binaries composed of neutron stars and/or black holes.
Such coalescence events are strong gravitational-wave emitters.
We demonstrate that measurements made by the LIGO and Virgo gravitational-wave observatories can be leveraged to severely constrain the validity of FRB binary coalescence models.
Existing measurements constrain the binary black hole rate to approximately $5\%$ of the FRB rate, and results from Advanced LIGO's O1 and O2 observing runs may place similarly strong constraints on the fraction of FRBs due to binary neutron star and neutron star--black hole progenitors.

\end{abstract}

\maketitle

%%%%%%%%%%%
\section{Introduction}
%%%%%%%%%%%

Recent years have seen the emergence of fast radio bursts (FRBs) as a new class of radio transient.
FRBs are characterized by millisecond durations, $\sim$Jansky flux densities, and dispersion measures (DMs) consistent with sources at gigaparsec (Gpc) distances.
Now observed with a growing number of instruments, including the
Parkes~\citep{Lorimer2007,Keane2011,Thornton2013,Burke-Spolaor2014,Petroff2014,Ravi2015,Champion2015,Keane2016},
Arecibo~\citep{Spitler2014,Spitler2016},
and Green Bank~\citep{Masui2015}
telescopes, FRBs are becoming increasingly accepted as a true astronomical phenomenon, rather than local signals of terrestrial origin.
Recently,~\cite{Keane2016} have even reported the first identification of an FRB's host galaxy, although this claim is currently disputed~\citep{Vedantham2016,Williams2016}.
FRBs also appear to be quite numerous.
While only \blue{17} FRBs have been reported to date (\citealt{Petroff2016}; see the \textsc{frbcat}\footnote{\url{http://www.astronomy.swin.edu.au/pulsar/frbcat/}}), after correcting for sky coverage and observing cadence it is estimated that between $10^3$ and $10^4$ occur on the sky per day~\citep{Thornton2013,Keane2015}.
That is, a hypothetical telescope array observing continuously with complete sky coverage would observe between 1000 and 10,000 FRBs per day.

A large number of theories have been put forward as to the possible source(s) of FRBs.
Theorized sources include (but are certainly not limited to) the collapse of
supramassive neutron stars~\citep{Falcke2014,Ravi2014,Zhang2014},
supergiant neutron star pulses~\citep{Connor2016,Cordes2015},
pulsar-planet systems~\citep{Mottez2014},
bremmstrahlung from gamma-ray bursts or active galactic nuclei~\citep{Romero2016},
and galactic flare stars~\citep{Loeb2014,Maoz2015}.
More exotic sources include the explosions of white holes~\citep{Barrau2014} and primordial black hole evaporation~\citep{Keane2012}.

Compact binary coalescences (CBCs) represent another broad class of theorized FRB progenitor.
The mergers of
binary neutron stars (BNS; \citealt{Totani2013,Wang2016a}),
neutron star--black hole (NSBH) binaries~\citep{Mingarelli2015},
white dwarf binaries~\citep{Kashiyama2013},
and charged binary black holes (BBHs; \citealt{Liu2016,Zhang2016})
have all been put forward as possible sources of FRB emission.
Furthermore, the recent localization of FRB\,150418 to an elliptical galaxy with low star formation would, if correct, support a CBC progenitor~\citep{Keane2016}.

The possibility that binary coalescences are FRB progenitors is particularly attractive.
If this were indeed the case, then FRBs would be promising electromagnetic counterparts to gravitational-wave detections of compact binary mergers, and would be valuable targets for future multi-messenger studies~\citep{Kaplan2015,Yancey2015}.
The recent discovery of the repeating fast radio burst FRB 121102~\citep{Spitler2016} points to a non-cataclysmic origin for at least some fraction of FRBs, ruling out binary coalescences as the sole progenitors of all FRBs.
As has been pointed out, though, FRBs may not constitute a single population~\citep{Mingarelli2015,Spitler2016};
there may instead exist multiple FRB populations, each arising from a different class of progenitor.
If binary coalescences are to be considered plausible models for one such progenitor population, then their astrophysical rates must be consistent with the inferred rate of FRBs.

In this Letter, we demonstrate that existing and future gravitational-wave measurements of the rates of binary coalescences can be leveraged to place novel constraints on the nature of FRB progenitors.
In some cases, we can confidently rule out certain classes of binary coalescences as dominant FRB progenitors.

%%%%%%%%%%%%%%%%%%%%%%%%%
\section{Rates of Compact Binary Coalescences}
\label{ratesSection}
%%%%%%%%%%%%%%%%%%%%%%%%%

The recent Advanced LIGO and Virgo detection of the BBH merger GW150914~\citep{LIGO2016a} has produced the first direct measurement of the binary black hole merger rate per comoving volume (the so-called ``rate density'') in the nearby Universe. From this event, it is inferred that the BBH merger rate density lies between $2$ and $400\rateUnits$~\citep{Abbott2016}.

While the rate densities of BNS and NSBH mergers remain unknown, binary pulsar observations and population synthesis models place rough bounds on the expected BNS and NSBH rates, respectively.
BNS and NSBH merger rate densities are predicted to plausibly fall between $R\bns=\blue{10-10^4\rateUnits}$ and $R\nsbh=\blue{0.6-10^3\rateUnits}$~\citep{Kalogera2004,OShaughnessy2008,Abadie2010,Kim2015}.
Note, however, that~\cite{Dominik2013} predict NSBH rate densities as low as $0.04\rateUnits$.
Gravitational-wave experiments have not yet begun to probe these predicted ranges;
the best experimental results, placed by jointly by Initial LIGO and Initial Virgo, limit BNS and NSBH merger rate densities to $R\bns<\blue{1.3\times10^5\rateUnits}$ and $R\nsbh<\blue{ 3.1\times10^4\rateUnits}$, respectively~\citep{Abadie2012a}.

Although the Initial LIGO/Virgo limits are well above the most optimistic predictions from population synthesis and binary pulsars, Advanced LIGO's recently concluded first observing run (O1) is expected to measure rate densities down to $R\bns\approx\blue{3\times10^3\rateUnits}$ and $R\nsbh\approx\blue{750\rateUnits}$, experimentally probing for the first time the range of astrophysically plausible merger rates~\citep{Abbott2016a}.
In 2017-18, Advanced LIGO's second observing run (O2) is projected to be sensitive to rate densities as low as $R\bns\approx\blue{450\rateUnits}$ and $R\nsbh\approx\blue{100\rateUnits}$, while the its third run (O3; 2018-19) further pushes Advanced LIGO's sensitivity to $R\bns\approx\blue{100\rateUnits}$ and $R\nsbh\approx\blue{20\rateUnits}$.

%%%%%%%%%%%%%%%%%%%%%%%%
\section{Rates of FRBs}
\label{frbRatesSection}
%%%%%%%%%%%%%%%%%%%%%%%%

The predicted and measured rates of binary coalescences allow for direct constraints on the nature of FRB progenitors by comparison to the inferred FRB rate per comoving volume.
Other authors have considered the physical rate of FRBs, but these calculations are typically not shown in detail and significant disagreement exists in the literature, e.g.~\cite{Totani2013} vs.~\cite{Zhang2016,Zhang2016b}.
Our goal in this section is therefore a careful accounting of the FRB rate density.
As we will show below, the FRB rate per comoving volume is potentially far higher than the corresponding rate densities of binary coalescences.
Thus, it is unlikely that the coalescence of stellar-mass compact binaries represents more than a small fraction of FRB progenitors.
Because of this rate discrepancy, the \textit{lowest} FRB rate estimates are \textit{most compatible} with CBC progenitors.
In the following, we will therefore deliberately seek a lower limit on the FRB rate density in order to most generously assess the plausibility of CBC progenitors of FRBs.

The inferred FRB rate per comoving volume is approximately $(3 r\obs)/(4\pi D^3)$.
Here, $D$ is the comoving distance containing the observed FRB population and $r\obs$ is the observed rate at which FRBs occur on the sky.
For simplicity, we will assume this rate density is constant and neglect evolution with redshift.
If FRB emission is \textit{beamed}, then the rate $r\obs$ is undercounted due to selection effects -- beamed FRBs, like pulsars or GRBs, are only observed if the Earth lies within the path of the beam.
In general, the FRB rate per comoving volume is
	\begin{equation}
	R\frb \approx \frac{3 r\obs}{\Omega D^3},
	\end{equation}
where $\Omega$ is a typical solid angle over which emission is beamed.

Although few FRBs have been observed, their inferred rate on the sky is large.
With four FRB detections at high Galactic latitudes using Parkes, \cite{Thornton2013} inferred that $r\obs=\blue{1.0^{+0.6}_{-0.5}\times10^4}$ FRBs occur on the sky per day.
However, there remains considerable disagreement as to the true value of $r\obs$, with subsequent radio surveys producing differing rate estimates, often defined with respect to different fluence limits and different assumptions about search systematics.

\cite{Keane2015}, for instance, point out that FRB detection is subject to significant selection effects, such as survey incompleteness below a fluence of $\sim2\,\text{Jy}\,\text{ms}$, suboptimal recovery of broad radio pulses, and potential obscuration of FRBs in the galactic plane.
They estimate a fluence-limited detectable FRB rate of $2500\skyRate$ above $\sim2\Jy\ms$.
\cite{Macquart2015} also arrive at $r\obs\approx2500\skyRate$ but by different means, suggesting that the apparent FRB rate at high latitudes is enhanced by interstellar scintillation.
\cite{Rane2015} adopt a Bayesian approach, combining several published rate estimates to obtain $r\obs = 4.4^{+5.2}_{-3.1}\times10^3\skyRate$ above $4.0\Jy\ms$.
On the other hand, \cite{Law2015} argue that previously published single-dish rate estimates are biased below their true values and that, once potential biases are corrected, previous estimates are consistent with $r\obs = 1.2\times10^4\skyRate$ above $1.7\Jy\ms$.

It is not obvious which value to select for $r\obs$ (or even which range of uncertainties to adopt).
In order to place a lower limit on the FRB rate density, however, we will take $r\obs = 2500\skyRate$, consistent with the lowest of the above estimates.
To additionally allow for various search selection effects, we will define $\eta$ as the FRB detection efficiency, the fraction of otherwise detectable FRBs (e.g. with intrinsic signal-to-noise ratios above some threshold detection value) which are actually recovered in a radio transient search.
The physical rate of FRBs on the sky is then $r\obs/(\eta\Omega)$.

\begin{figure}
  \centering
  \includegraphics[width=.48\textwidth]{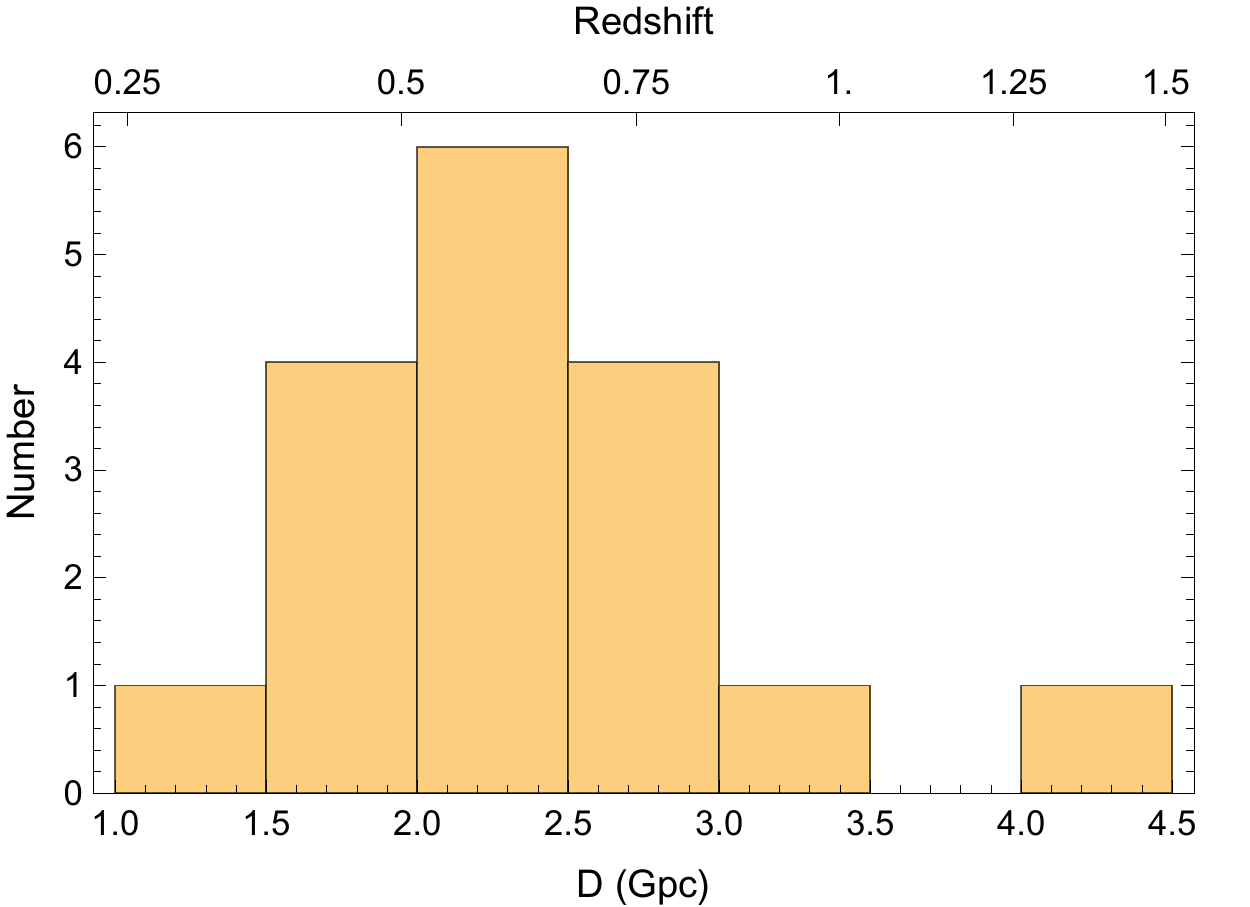}
  \caption{
  Distribution of inferred distances to known FRBs, assuming a homogeneous, fully ionized intergalactic medium and neglecting dispersion measure contributions from both the Milky Way and FRB host galaxies.
  We take $3\Gpc$ as a fiducial distance bounding the observed FRB population.
  }
  \label{distPlot}
\end{figure}

Distances to FRB sources may be estimated using their reported DMs, which we obtained from the \textsc{frbcat}~\citep{Petroff2016}.
Assuming that the intergalactic medium (IGM) is homogeneous and fully ionized, the dispersion measure $\text{DM}_\textsc{igm}$ due to propagation through the IGM is related to source redshift via~\citep{Ioka2003,Inoue2004}
\begin{equation}
	\label{DM}
	\text{DM}_\textsc{igm}(z) = \frac{\overline{n}_e c}{H_0}\int_0^z \frac{(1+z')\,dz'}{\sqrt{\Omega_\textsc{m}(1+z')^3 + \Omega_\Lambda}},
\end{equation}
where $\overline n_e = \rho_c \Omega_\textsc{b} / m_p = 2.5\times10^{-7}\,\text{cm}^{-3}$ is the local free electron density in a fully ionized Universe.
Here, $\Omega_\textsc{b}$, $\Omega_\textsc{m}$, and $\Omega_\Lambda$ are the energy densities of baryons, matter, and dark energy, respectively, $m_p$ is the proton mass, and $\rho_c = 3H_0^2/8\pi G$ is the critical energy density required to close the Universe.
$G$ is Newton's constant, $c$ the speed of light, and $H_0$ the Hubble constant;
we use the \cite{PlanckCollaboration2015} parameters $H_0 = 67.7\,\text{km}\,\text{s}^{-1}\text{Mpc}^{-1}$, $\Omega_\textsc{b} = 0.049$, $\Omega_\textsc{m} = 0.31$, and $\Omega_\Lambda = 0.69$.
In the small redshift limit, Eq.~\eqref{DM} reduces to $\text{DM}_\textsc{igm}(z) \approx \overline n_e D$, where $D= c z/H_0$ is the approximate source distance at low redshifts.
In general, comoving distance is given by
\begin{equation}
	\label{r}
	D(z) = \frac{c}{H_0} \int_0^z \frac{dz'}{\sqrt{\Omega_\textsc{m}(1+z')^3 + \Omega_\Lambda}}.
\end{equation}
Using Eqs.~\eqref{DM} and \eqref{r}, the inferred comoving distances to the 17 known FRBs are shown in Fig.~\ref{distPlot}.
Based on this sample, we will take $D=\blue{3\Gpc}$ as a fiducial distance encompassing the observed FRB population.

We have made several assumptions in computing the distances shown in Fig.~\ref{distPlot}.
Since a factor of $2$ error in the fiducial distance will result in a factor of $2^3$ error in the FRB rate density, it is important to highlight these assumptions and understand how they affect our result.
First, we assumed that the observed radio dispersions are entirely due to propagation through the IGM.
In reality, the Milky Way may contribute up to $\sim20\%$ of the observed DM~\citep{Petroff2016}.
The distances in Fig.~\ref{distPlot} may therefore be overestimated by a factor of $\sim1.25$.
If we also allow for a comparable DM contribution by the FRB's host galaxy (as well contributions from any matter overdensities along the line of sight to the FRB), then the distances may be overestimated by at least a factor of $1.7$.
This implies that our FRB rate density is \textit{underestimated} by a factor between $2$ and $5$.

Second, we assumed a fully ionized Universe.
While valid for hydrogen, this is not necessarily true for helium, which may be either singly or fully ionized.
Helium makes up approximately $24\%$ of the IGM by mass~\citep{Inoue2004}; if this helium is only singly ionized, then the free electron density $\overline n_e$ will be reduced by roughly $10\%$.
Finally, $\Omega_\textsc{b}$ is an overestimate of the baryon density in the IGM, since $\sim10\%$ of baryons are sequestered in galaxies~\citep{Fukugita2004}.
Together, these two approximations cause $\overline n_e$ to be (at most) $20\%$ larger than the true free electron density in the IGM.
By overestimating $\overline n_e$, the fiducial distance $D$ is underestimated by a factor of $1.25$, and the FRB rate density is overestimated by a factor of $2$.

Note that, of the assumptions described here, the first (uncertainty in the intergalactic DM) will cause the fiducial distance to be underestimated, while the second and third (uncertainty in $\overline n_e$) cause the distance to be overestimated.
Of these uncertainties, the potentially large overestimate of the intergalactic DM is expected to dominate.
Thus our choice of $D=3\Gpc$ is likely an upper bound on the fiducial FRB distance.
Because $R\frb \propto D^{-3}$, any decrease in the fiducial distance will only increase the FRB rate density, further increasing the tension between the rates of FRBs and binary coalescences.

All together, the FRB rate per comoving volume is
	\begin{equation}
	\begin{aligned}
	R\frb = &\left(8.1\times10^3\,\text{Gpc}^{-3}\,\text{yr}^{-1}\right) \left(\frac{r\obs}{2500\,\text{sky}^{-1}\,\text{day}^{-1}}\right) \\
		&\times \left(\frac{1}{\eta}\right)\left(\frac{3\,\text{Gpc}}{D}\right)^3\left(\frac{4\pi\,\text{Sr}}{\Omega}\right).
	\end{aligned}
	\end{equation}	
The lowest plausible FRB rate density, corresponding to $r\obs = 2500\,\text{sky}^{-1}\,\text{day}^{-1}$, radio transient detection efficiencies of $\eta=1$, and \textit{isotropic} radio emission ($\Omega=4\pi\,\text{Sr}$), is
	\begin{equation}
	R\frb^\text{Low} = 8.1\times10^3\rateUnits.
	\label{lowRate}
	\end{equation}
A more realistic rate density, on the other hand, is obtained by assuming $r\obs=5000\skyRate$, a detection efficiency of $\eta = 0.5$, and beamed emission with a $30^\circ$ half-opening angle.
These values give
	\begin{equation}
	R\frb^\text{Realistic} = 4.8\times10^5\rateUnits,
	\label{realRate}
	\end{equation}
nearly two orders of magnitude larger than Eq.~\eqref{lowRate}.

Our lower limit on the FRB rate density agrees well with the rate previously estimated by~\cite{Totani2013}.
It is, however, more than an order of magnitude higher than the more recent result computed by~\cite{Zhang2016,Zhang2016b}.
The discrepancy lies in the fact that~\cite{Zhang2016,Zhang2016b} improperly uses the luminosity distance $D_L = D(1+z)$ rather than the comoving distance to calculate the FRB rate density.
\cite{Zhang2016,Zhang2016b} chooses $z=1$ as a fiducial redshift, at which $D = 3.4\Gpc$ while $D_L = 6.8\Gpc$.
This factor of $2$ adjustment in distance leads to a factor of $8$ discrepancy in the FRB rate density.
Using $D=3.4\Gpc$ in Eq.~(8) of~\cite{Zhang2016b} gives an FRB rate density of $5.8\times10^3\rateUnits$, in reasonably good agreement with our lower limit.

%%%%%%%%%%%%%%%%%%%%%%%%
\section{Compact Binaries as FRB Progenitors?}
\label{fractionSection}
%%%%%%%%%%%%%%%%%%%%%%%%
	
\begin{figure}
  \centering
  \includegraphics[width=.48\textwidth]{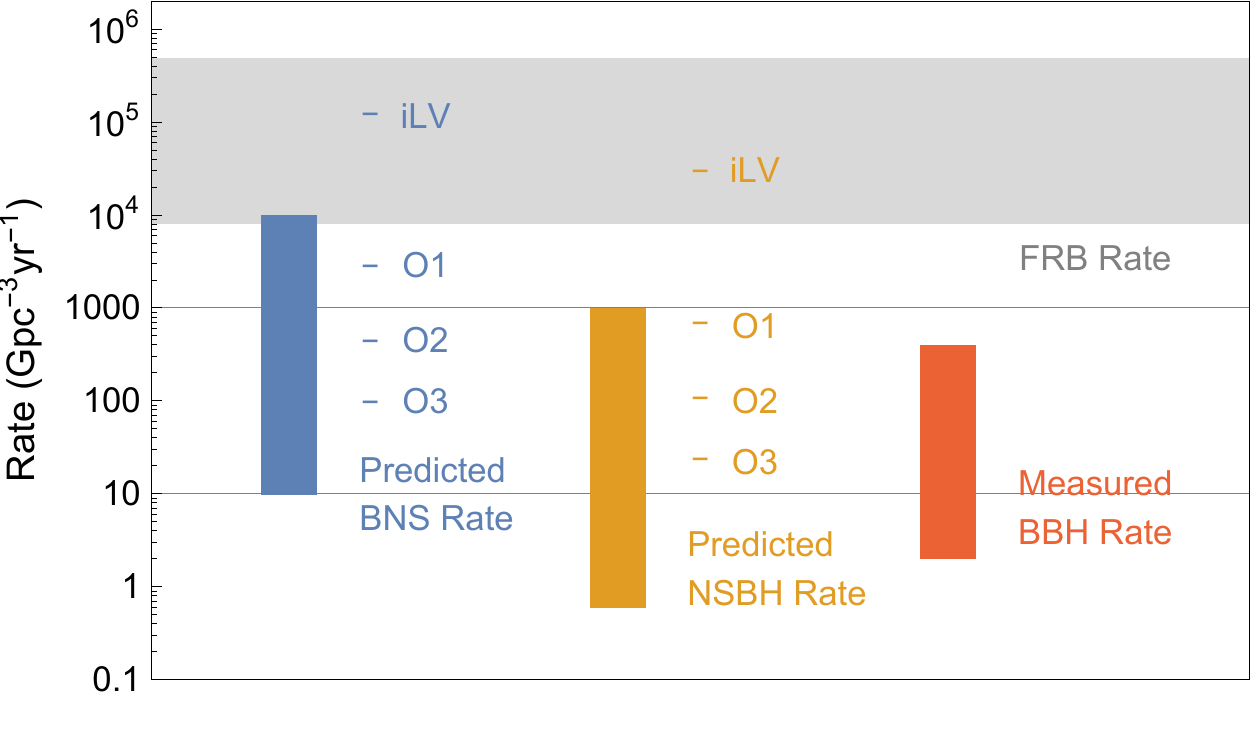}
  \caption{
  Binary coalescence rates compared to the inferred rate of FRBs.
  Solid bars indicate the range of BNS (blue) and NSBH (orange) merger rates predicted by binary pulsar observations and population synthesis models~\citep{Abadie2010}, as well as the \textit{measured} LIGO/Virgo rate of BBH mergers (red; \citealt{Abbott2016}).
  Also shown are existing Initial LIGO/Virgo (iLV) limits and projected O1, O2, and O3 sensitivities~\citep{Abadie2012a,Abbott2016a}.
  The gray band indicates a range of potential FRB rate densities, from the lowest plausible value in Eq.~\eqref{lowRate} to a more realistic estimate in Eq.~\eqref{realRate}.
  }
  \label{fracPlots}
\end{figure}

By comparing $R\frb$ from Sect.~\ref{frbRatesSection} to the binary coalescence rates in Sect.~\ref{ratesSection}, we can constrain the fraction of FRBs that can be explained via compact binary coalescences.
Fig.~\ref{fracPlots} shows a range of potential FRB rate densities, from the lowest plausible estimate given in Eq.~\eqref{lowRate} (assuming $r\obs=2500\skyRate$, efficiency $\eta=1$, and isotropic FRB emission) to the more realistic value in Eq.~\eqref{realRate} (which assumes $r\obs=5000\skyRate$, efficiency $\eta=0.5$, and FRB beaming with a half-opening angle of $30^\circ$).
Solid bars indicate the range of BNS and NSBH merger rate densities predicted by binary pulsar observations and population synthesis models, as well as the measured BBH rate density.
Also shown are existing Initial LIGO/Virgo limits, as well as the projected sensitivities of the O1, O2, and O3 observing runs.

\textit{Binary black holes:}
the measured rate of binary black holes mergers is \textit{at most} $\blue{\sim5\%}$ of the inferred FRB rate.
Thus, BBHs cannot explain more than a small fraction of the observed FRB population.
Previous claims that the rates of FRBs and BBH mergers are consistent~\citep{Zhang2016,Zhang2016b} are based on an erroneous calculation of the FRB rate density, as discussed in Sect.~\ref{frbRatesSection}.

\textit{Neutron star-black hole binaries:}
population synthesis predictions are highly inconsistent with the theory that NSBH mergers are FRB progenitors, with predicted NSBH merger rates equal to at most $\blue{\sim12\%}$ of the FRB rate.
This fraction assumes isotropic radio emission, and hence should be taken as a highly optimistic upper limit on the FRB fraction compatible with NSBH binaries.
Even moderate beaming, with a half-opening angle of e.g. $30^\circ$, reduces the predicted FRB fraction to $\blue{\sim0.8\%}$.
Assuming the realistic FRB rate density in Eq.~\eqref{realRate} further lowers this fraction by a factor of four.

Although Initial LIGO/Virgo upper limits are uninformative (limiting the most optimistic NSBH fraction of FRB progenitors to $\blue{R\nsbh/R^\text{Low}\frb\lesssim4}$), Advanced LIGO is capable of measuring significantly smaller NSBH merger rates.
A non-detection during the O1 and O2 observing runs, for instance, would limit the NSBH FRB fraction to $\blue{\lesssim9\%}$ and $\blue{\lesssim1\%}$, respectively (assuming isotropic emission).

\textit{Binary neutron stars:}
there exist competing claims as to whether the rates of FRBs and binary neutron star mergers are~\citep{Totani2013,Wang2016a} or are not~\citep{Thornton2013} compatible.\footnote{
Note that \cite{Wang2016a} adopt the FRB rate estimate of \cite{Zhang2016b}, and hence incorrectly find strong consistency between the rates of BNS mergers and FRBs.}
We find that the most optimistic BNS rate density predictions are consistent with the lowest possible FRB rate density, with $\blue{R\bns/R^\text{low}\frb\approx1.2}$ 
Therefore, BNS mergers could constitute a subpopulation of FRB progenitor if multiple FRB subclasses do indeed exist.
This compatibility is tenuous, however, simultaneously requiring the highest possible BNS rates and the lowest possible FRB rates (with, e.g., perfect FRB detection efficiency and isotropic radio emission).
FRB models that predict even moderately beamed emission are largely incompatible with BNS progenitors.

If BNS mergers are indeed FRB progenitors, then it is likely that Advanced LIGO will observe a large number of BNS sources in the O1 observing run.
If no such detections are made, then the resulting rate limits will increasingly cast doubt on the role of BNSs as FRB progenitors.
An Advanced LIGO non-detection during O1 and O2 would limit the most optimistic fraction of FRBs compatible with BNS mergers to $\blue{\lesssim40\%}$ and $\blue{\lesssim6\%}$, respectively.
If we assume moderate FRB beaming (again with a half-opening angle of $30^\circ$), then O1 and O2 non-detections imply even more stringent FRB fractions of $\blue{\lesssim2\%}$ and $\blue{\lesssim0.4\%}$, respectively.
Note that these limits also apply equally to short-lived products of BNS mergers, such as hypermassive neutron stars.

%%%%%%%%%%%
\section{Conclusions}
%%%%%%%%%%%

A diverse range of FRB progenitor models have been proposed, including the binary coalescences of neutron stars and/or black holes.
Existing or future limits from gravitational-wave observations can serve to severely constrain such models.
The recent Advanced LIGO/Virgo measurement of the local BBH merger rate density largely rules out stellar-mass binary black holes as progenitors of the observed FRB population.
Meanwhile, predictions of NSBH merger rate densities from population synthesis are in strong tension with the inferred rate density of FRBs;
upcoming observations by Advanced LIGO and Virgo could rule out NSBHs as FRB progenitors.

Under generous assumptions (broadly beamed radio emission, large FRB distances, and low underlying FRB rates), the rate of BNS mergers may be consistent with a subpopulation of FRB progenitors.
In order for this subpopulation to be significant, however, the BNS merger rate density must be on the order of $\sim10^4\rateUnits$, comparable to the most optimistic predictions from population synthesis.
Additionally, FRB emission must be largely isotropic; models that predict even moderately beamed emission are inconsistent with BNS rates.
If BNS mergers are indeed FRB progenitors, then Advanced LIGO and Virgo will soon begin to observe a large number of such systems.
If no such observations are made, then the resulting rate limits will increasingly constrain the ability of BNSs to explain even a subclass of the FRB population.

\acknowledgements
We thank Krzysztof Belczynski, Christopher Berry, Daniel Hoak, Brennan Hughey, Shri Kulkarni, and Graham Woan, as well as the anonymous referee whose comments greatly improved this text.
The authors are members of the LIGO Laboratory, supported by the United States National Science Foundation.
LIGO was constructed by the California Institute of Technology and Massachusetts Institute of Technology with funding from the National Science Foundation and operates under cooperative agreement PHY-0757058.
This paper carries LIGO Document Number P1600106.
A Mathematica notebook that reproduces the results in this paper is available at: \url{https://github.com/tcallister/GravWaveFRBConstraints}.

\software{Mathematica}

\end{document}